\documentstyle [twocolumn,epsf,amssymb]{mn}
\newcommand{\mincir}{\raise
-3.truept\hbox{\rlap{\hbox{$\sim$}}\raise4.truept\hbox{$<$}\ }}
\newcommand{\magcir}{\raise
-3.truept\hbox{\rlap{\hbox{$\sim$}}\raise4.truept\hbox{$>$}\ }}
\newcommand{\minmag}{\raise
-3.truept\hbox{\rlap{\hbox{$<$}}\raise5.truept\hbox{$<$}\ }}
\newcommand{\be}{\begin{equation}}
\newcommand{\ee}{\end{equation}}
\newcommand{\CC}{\Lambda}
\newcommand{\rLo}{\rho_{\CC}^0}
\newcommand{\rL}{\rho_{\CC}}
\newcommand{\rmr}{\rho_m}

\newcommand{\wm}{\omega_m}
\newcommand{\wD}{\omega_{\rm D}}
\newcommand{\rc}{\rho_c}
\newcommand{\rco}{\rho_{c}^0}
\newcommand{\rmo}{\rho_{m}^0}

\newcommand{\fM}{f_{m}}
\newcommand{\fL}{f_{\Lambda}}
\newcommand{\Om}{\Omega_m}
\newcommand{\OL}{\Omega_{\Lambda}}

\newcommand{\tOm}{\tilde{\Omega}_m}

\newcommand{\tOD}{\tilde{\Omega}_{\rm D}}
\newcommand{\rD}{\rho_{\rm D}}

\newcommand{\xiM}{\xi}
\newcommand{\xiR}{\xi'}
\newcommand{\rRo}{\rho_r^0}
\newcommand{\Omo}{\Omega_m^0}

\newcommand{\ORo}{\Omega_{r}^0}
\newcommand{\OLo}{\Omega_{\Lambda}^0}
\newcommand{\ba}{\begin{eqnarray}}
\newcommand{\ea}{\end{eqnarray}}
\newcommand{\brr}{\begin{array}}
\newcommand{\err}{\end{array}}
\newcommand{\bc}{\begin{center}}
\newcommand{\ec}{\end{center}}

\renewcommand{\baselinestretch}{1.0}

\title[Effective equation of state for running vacuum]
{Effective equation of state for running vacuum: ``mirage'' quintessence
and phantom dark energy}
\author[Spyros Basilakos \& Joan Sol\`a]{
Spyros Basilakos$^1$ and Joan Sol\`a$^{2}$\\
\vspace{0.1cm} $^1$ Academy of Athens, Research Center for Astronomy \&
Applied
  Mathematics, Soranou Efessiou 4, 11-527, Athens, Greece\\
$^2$ High Energy Physics Group, Dept. ECM, and Institut de
Ci\`encies del Cosmos (ICC), Universitat de Barcelona, Av. Diagonal 647\\
\vspace{0.1cm}
 E-08028 Barcelona, Catalonia, Spain\\ E-mails: svasil@academyofathens.gr, sola@ecm.ub.edu,
}

\begin{document}

\maketitle

\begin{abstract}
{Past analyses of the equation of state (EoS) of the Dark Energy
(DE) were not incompatible with a phantom phase near our time}. This has
been the case in the years of WMAP observations, in combination with the
remaining cosmological observables. {Such situation did not
completely disappear from the data collected from the {\it Planck}  satellite
mission}. {In it the EoS analysis may still be interpreted as
suggesting $\wD\lesssim-1$}, and so a mildly evolving DE cannot be
discarded. In our opinion the usual ansatzs made on the structure of the
EoS for dynamical DE models (e.g. quintessence and the like) are too
simplified. In this work we examine in detail some of these issues and
suggest that a general class of models with a dynamical vacuum energy
density could explain the persistent phantom anomaly, despite there is no
trace of real phantom behavior in them. The spurious or ``mirage'' effect
is caused by an attempt to describe them as if the DE would be caused by
fundamental phantom scalar fields. Remarkably, the effective DE behavior
can also appear as quintessence in transit to phantom, or vice versa.

\vskip0.2cm

\end{abstract}

\section{Introduction}

Dark energy is one of the most mysterious components of our universe
(Huterer \& Turner, 1999). Despite very little is known on its ultimate
nature, compelling evidence exists from independent data sets derived from
the luminosity-redshift relation of distant supernovae, the anisotropies
of the CMB, the BAO scale produced in the last scattering surface by the
competition between the pressure of the coupled baryon-photon fluid and
gravity, and the matter power spectrum obtained from the large scale
structures (LSS) of the Universe. All of them speak up convincingly for
the existence of such mysterious and overwhelmingly abundant entity  (cf.
Riess et al. 2004; Knop et al. 2004; Spergel et al. 2007; Blaket et al.
2011; Amanullah et al. 2010; Komatsu et al. 2011; and references therein).
The recently released analysis of the {\it Planck}  satellite data on the
anisotropies of the CMB reinforce the previous observations and confirm
that the DE component is an indispensable ingredient of the cosmological
puzzle (Ade et al. 2013). {Let us however note that CMB data alone
are not critically sensitive to DE unless we combine them with other data
from large scale structure or cosmic distance measurements. The existence
of DE, first directly observed by supernova measurements (Riess et al,
1998; Perlmutter et al, 1999), is required (Spergel et al., 2003) by the
combination of CMB power spectrum measurements and any one of the
following low redshift observations: measurements of the Hubble constant,
measurements of the galaxy power spectrum, galaxy cluster abundances, or
supernova measurements of the redshift-distance relation. So the main
constraints on expansion history of the universe and reconstructed
equation of state of DE is not coming from just CMB data. This is due to
the ``geometric degeneracy'' which prevents both the curvature and
expansion rate from being determined simultaneously by the CMB alone
(Bond, Efstathiou \& Tegmark, 1997; Zaldarriaga, Spergel, \& Seljak,
1997). However, inclusion of CMB lensing power spectrum data, which probe
structure formation and geometry long after decoupling  breaks the CMB
geometric degeneracy (B. D. Sherwin et al, 2011; Hazra, Shafieloo \&
Souradeep, 2013).}

Popular proposals for the DE are, among others, quintessence and phantom
energy in its various forms, modified gravity etc. (cf. Peebles \& Ratra
2003; Padmanabhan 2003; Lima 2004). Interestingly enough, the {\it Planck}
 observations (Ade et al. 2013) suggest an effective equation of state
(EoS) for the DE centered in the phantom domain $\wD\lesssim -1$,
specifically in the range $\wD=-1.13^{+0.13}_{-0.10}$. This situation is
not new, as it had also been reckoned by the long series of WMAP
observations (Spergel et al. 2007; Komatsu et al. 2011) and indeed it has
triggered many works in the literature (see e.g. Alam, Sahni \&
Starobinsky 2004; Jassal, Bagla \& Padmanabhan 2005 and 2010; Feng, Wang
\& Zhang 2005; Caldwell \& Doran 2005; Sol\`a \& \v{S}tefan\v{c}i\'{c}
2005 and 2006).

The observed persistence of this result in the hot data from the {\it
Planck}  mission (Ade et al. 2013) is still not compelling evidence of a
phantom phase, but depending on the combination of this data with other
sources yields a result that is in tension with the $\wD=-1$ standard
expectation at more than $2\sigma$ within the phantom domain. This could
be interpreted again as a symptom that the DE might have a mild dynamical
behavior, as we do not expect a constant EoS result $\wD<-1$, which would
be difficult to expalin. Such dynamics should be very desirable since the
idea of a rigid cosmological constant (CC) or vacuum energy is very
difficult to reconcile with a possible solution of the cosmological
constant problems plaguing theoretical cosmology (Weinberg 1989;
Padmanabhan 2003). A constant CC term throughout the entire history of the
universe presents strong conceptual difficulties from the point of view of
fundamental physics. The apparent lack of evidence of the dynamical
component of the DE in the observations could be attributed, in our view,
to the fact that it has been tested using exceedingly simple
parametrizations which might be blind to this kind of effects. We cannot
exclude that the effects are too small, of course, but what is most
encouraging is that from the theoretical point of view we expect a
dynamical behavior of the vacuum energy in quantum field theory (QFT) in
curved spacetime whose evolution should be accessible to future
phenomenological observations (for reviews see e.g. Sol\`a 2011, 2013). We
will show some examples in this paper.

There are a number of attempts in the literature proposing a dynamical
vacuum energy in various ways (cf. e.g. Ozer \& Taha 1986; Peccei, Sol\`a
\& Wetterich 1987; Freese et al. 1987; Peebles \& Ratra 1988;  Wetterich
1988; Carvalho, Lima \& Waga 1992, see also the review by Overduin \&
Cooperstock, 1998 and references therein for the old literature). However,
a deeper theoretical approach can be obtained from the point of view of
the so-called running vacuum energy in QFT, where the CC is expected to
show some dynamical evolution with the expansion history of the universe
(Shapiro \& Sol\`a 2002, 2003, 2004; Babi{\'c} \textit{et al.} 2002;
Sol\`a 2007; Shapiro \& Sol\`a 2009). This idea has been tested
successfully in recent studies using the available data on expansion and
structure formation (cf. Basilakos, Plionis \& Sol\`a 2009; Grande et al.
2011; Basilakos, Polarski \& Sol\`a 2012). Such framework could also help
to shed some light on the cosmic coincidence problem (Grande, Sol\`a \&
\v{S}tefan\v{c}i\'{c}, 2006, 2007; Grande, Pelinson \& Sol\`a, 2008).
Furthermore, very recently a class of models of this sort has been
employed to describe the full cosmological history from inflation to the
present time (cf. Lima, Basilakos \& Sol\`a 2012; Perico et al. 2013,
Basilakos, Lima \& Sol\`a 2013).

In this paper we wish to dwell on a wide class of dynamical models of the
vacuum energy inspired in QFT in curved spacetime in which the effective
EoS appears indeed as evolving with the expansion of the universe, i.e. it
is seen as a slow function of the redshift, $\wD=\wD(z)$. We will show
that $\wD(z)$ can approach very close to $-1$ near our time (even take
this exact value at present), both from above and from below. In the first
case the dynamical vacuum model mimics quintessence, and in the second
phantom energy. The theoretical vacuum model, however, does not hinge on
the existence of fundamental quintessence or phantom fields. It only
assumes the QFT running of $\rL$ and $G$ caused by the matter effects in
QFT in curved spacetime. Such ``mirage'' behavior of the EoS is caused by
our modeling of the dynamical vacuum energy as if it were a dynamical
scalar field. This kind of feature could give some clue as to the
interpretation of the phantom phase that persistently appeared in the
analysis of the WMAP and stays in the recent {\it Planck}  data.

The structure of the paper is as follows. In section 2 we discuss the
cosmologies with running $\rL$ and/or $G$. In section 3, we introduce the
general form of effective equation of state for the dark energy in these
cosmologies. In section 4 we compute in detail the effective equation of
state for specific models of this kind. The corresponding numerical
analysis is presented in section 5. The final discussion and main
conclusions are summarized in section 6.

\section{Expansion dynamics with running parameters}\label{sect:dynamicalvac}

When modeling the expanding universe as a perfect fluid with
matter-radiation density $\rho_{m}$, and corresponding pressure
$p_{m}=\omega_{m} \rho_{m}$\footnote{Here $\omega_{m}=0$ for
non-relativistic matter, and $\omega_{m}=1/3$ for the radiation
component (relativistic matter). The latter is negligible deep in
the matter dominated era. We will not consider transient situations
that interpolate between these two clearly differentiated epochs.},
the energy momentum tensor of matter reads ${T}_{\mu\nu}=
-p_{m}\,g_{\mu\nu}+(\rho_{m}+p_{m})U_{\mu}U_{\nu}$. It is well known
that the cosmological constant term $\Lambda\,g_{\mu\nu}$ on the
\textit{l.h.s.} of Einstein's equations can be absorbed on the
\textit{r.h.s.} with a modified energy-momentum tensor
$\tilde{T}_{\mu\nu}\equiv T_{\mu\nu}+g_{\mu\nu}\,\rL $, in which
$\rho_{\Lambda}=\Lambda/(8\pi G)$ is the vacuum energy density
associated to the presence of $\CC$, and the corresponding pressure
is $p_{\Lambda}=-\rho_{\Lambda}$. The field equations then read
formally the same way:
\begin{equation}
R_{\mu \nu }-\frac{1}{2}g_{\mu \nu }R=8\pi G\ \tilde{T}_{\mu\nu}\,,
\label{EE}
\end{equation}
in which the full energy momentum tensor including the effect of the
vacuum energy density takes also the same form:
$\tilde{T}_{\mu\nu}=-p_{\rm tot}\,g_{\mu\nu}+(\rho_{\rm tot}+p_{\rm
tot})U_{\mu}U_{\nu}$, with $\rho_{\rm tot}=\rho_{m}+\rho_{\Lambda}$ and
$p_{\rm tot}=p_{m}+p_{\Lambda}=p_{m}-\rho_{\Lambda}$, or explicitly,
\begin{equation}
\tilde{T}_{\mu\nu}= (\rho_{\Lambda}-p_{m})\,g_{\mu\nu}+(\rho_{m}+p_{m})U_{\mu}U_{\nu}\,.
\label{Tmunuideal}
\end{equation}
In the flat FLRW metric, the two independent gravitational field equations
are the following: \be
8\pi G (\rmr+\rL) = 3H^2 \;, \label{friedr} \ee
\be
 8\pi G (\wm\rmr-\rL) =-2{\dot H}-3H^2\;, \label{friedr2} \ee where
the overdot denotes derivative with respect to cosmic time $t$.

Although a rigid CC term is the simplest possibility, it is remarkable
that the Cosmological Principle embodied in the FLRW metric still permits
the vacuum energy to be a function of the cosmic time, $\rL=\rL(t)$ or of
any collection of homogeneous and isotropic dynamical variables
$\chi_i=\chi_i(t)$, i.e. $\rL=\rL(\chi_i(t))$. The same applies to the
gravitational coupling, which can also be a function of these dynamical
variables: $G=G (\chi_i(t))$. In the following, although we will usually
write $\rL=\rL(t)$ and $G=G(t)$ when these parameters evolve with time, we
shall indeed understand that they may depend on one or more fundamental
dynamical variable (for specific models, cf. Sect. \ref{sect:runningvac}).

Let us consider now the different conservation laws that emerge from a
FLRW-like cosmology with running parameters (cf. Sol\`a 2011, 2013 and
references therein). Despite the possible time evolution of the vacuum
energy, the corresponding EoS can still maintain the usual form
$p_{\Lambda}(t)=-\rho_{\Lambda}(t)$. The Bianchi identities, in their
turn, imply that the covariant derivative of the full \textit{r.h.s.} of
Einstein's equations (\ref{EE}) is zero, namely
$\bigtriangledown^{\mu}\,\left(G\,\tilde{T}_{\mu\nu}\right)=0$, where $G$
is also involved since we admit it could be variable. With the help of the
FLRW metric, it is easy to see that the generalized conservation law can
be cast as
\begin{equation}\label{BianchiGeneral}
\frac{d}{dt}\,\left[G(\rmr+\rL)\right]+3\,G\,H\,(1+\wm)\rmr=0\,.
\end{equation}
The concordance model, or  $\CC$CDM model (Peebles 1984), appears as
a particular case of that relation in which both $\rL$ and $G$ are
constant. In this case, the conservation law boils down to the
standard one, $\dot{\rho}_m+3\,H\,(1+\wm)\rmr=0$, whose solution in
terms of the scale factor is well-known:
\begin{equation}\label{solstandardconserv}
\rmr(a)=\rmo\,a^{-3(1+\wm)}\,.
\end{equation}
However, the identity (\ref{BianchiGeneral}) allows for interesting
possible generalizations of the local conservation laws, that lead to the
following three types of generalized models:

Type i) $\CC_t$CDM model: $\CC=\CC(t)$ variable, and $G=$const. In this
case, Eq.\,(\ref{BianchiGeneral}) gives
\begin{equation}
\dot{\rho}_{m}+3(1+\omega_{m})H\rho_{m}=-\dot{\rho_{\Lambda}}\,. \label{frie33}
\end{equation}
The assumption
$\dot{\rho_{\Lambda}}\neq 0$ necessarily requires some energy exchange
between matter and vacuum, e.g. through vacuum decay into matter, or vice
versa.

Type ii) $\CC_tG_t$CDM model: $\CC=\CC(t)$ variable, and $G=G(t)$ also
variable, assuming matter conservation (\ref{solstandardconserv}). In the
present instance the generalized conservation law amounts to
\begin{equation}\label{Bianchi1}
(\rmr+\rL)\dot{G}+G\dot{\rL}=0\,.
\end{equation}
Here the time evolution of the vacuum energy density is possible at the
expense of a running gravitational coupling. This is what permits local
covariant matter conservation in this case.

Type iii) $\CC G_t$CDM model: $\CC=$const. and $G=G(t)$ variable. The
generalized conservation law (\ref{BianchiGeneral}) renders
\begin{equation}\label{dGneqo}
\dot{G}(\rmr+\rL)+G[\dot{\rho}_m+3H(1+\wm)\rmr]=0\,.
\end{equation}
Here matter is not conserved, and the gravitational coupling is again
running. Despite the vacuum energy here is constant, this situation can
also produce an effective or ``mirage'' dynamical dark energy effect, as
we shall see and analyze in detail. For this reason we encompass this
model also as a dynamical DE model.

The independent variables of our cosmological analysis can be taken as
$\rmr$, $\rL$ and $H$. Needless to say none of the previous models can be
explicitly solved until (based on some theoretical motivation) a dynamical
law is given for one of the remaining cosmological variables, e.g. the
function $\rL=\rL(\chi_i(t))$ -- in the case of the $\CC_t$CDM and
$\CC_tG_t$CDM models -- or, in the alternative $\CC G_t$CDM model, a
matter density $\rmr$ that deviates from the standard conservation law
(\ref{solstandardconserv}). Assuming that $\wm$ is known, the remaining
two variables in the triad $(\rmr, \rL, H)$ can then be solved using
either the two Friedmann's equations (\ref{friedr})-(\ref{friedr2}), or
one of them together with the corresponding generalized conservation law
(\ref{frie33}), (\ref{Bianchi1}) or (\ref{dGneqo}). With this information,
and following the aforementioned procedure, the corresponding  Hubble
function and energy densities can be obtained for any of the models i),
ii) and iii) mentioned above (cf. Sect. \ref{sect:runningvac} for the
details).

We emphasize that the above three generalized models can stay sufficiently
close to the standard $\CC$CDM model provided the possible time variation
of $\rL$ and/or $G$ is sufficiently mild, and of course within bounds.
Finally, let us mention that the vacuum can run at fixed $G$ with local
covariant conservation of matter. In this situation, however, the DE must
be composite, i.e. there must be another DE component apart from $\CC$--
see e.g. the so-called $\CC$XCDM framework (Grande, Sol\`a \&
\v{S}tefan\v{c}i\'{c} 2006).

Because of the proximity of the above running models with the concordance
model, it is natural to ask what is the effective equation of state of the
dark energy for them, as perceived from an analysis in which the
cosmological parameters $\CC$ and $G$ are assumed strictly constant, in
particular $\CC=0$, and where the DE is attributed to some smoothly
evolving, self-conserved, dynamical entity, such as for example a
dynamical scalar field. In the next section we provide an answer to this
question.

\section{Effective equation of state for general running models}\label{sect:effectiveEOS}

Let us consider a generalized vacuum framework in flat space with
normalized Hubble function given as follows:
\begin{eqnarray}\label{eq:Hgeneral}
E^2(z)\equiv\frac{H^2(z)}{H_0^2}=\frac{8\pi\, G(z)}{3 H_0^2}\left[\rmr(z)+\rL(z)\right]\,,
\end{eqnarray}
in which we have explicitly accounted for a possible cosmic evolution of
both $G$ and $\rL$ as a function of the redshift $z$. It is convenient to
parametrize the above expression in the following way:
\begin{eqnarray}\label{eq:Hgeneral2}
E^2(z)=\Om^0\,\fM(z;r_i)(1+z)^{\alpha_m}+\OL^0\,\fL(z;r_i)\,,
\end{eqnarray}
with $\alpha_m\equiv3\,(1+\wm)$, and we have defined the current
cosmological parameters $\Om^0=\rmr^0/\rc^0$ and $\OL^0=\rL^0/\rc^0$
(superscript $0$ meaning at redshift $z=0$, i.e. now). Any of the model
types introduced in the previous section can eventually be brought to the
form (\ref{eq:Hgeneral2}). The two functions $\fM(z;r_i)$ and $\fL(z;r_i)$
of the redshift are assumed to be known in the given vacuum model (or at
least calculable), and may depend on some free parameters $r_i=r_1,r_2,..$
(Sol\`a \& \v{S}tefan\v{c}i\'{c} 2005 and 2006). Their explicit form will
depend on the possible evolution of the vacuum energy density $\rL$ and/or
of the gravitational coupling $G$ as functions of $z$, as well as on a
possible anomalous law for the matter energy density (specific examples
will be given in the next section). For the standard $\CC$CDM, these
functions take, of course, the trivial values: $\fM=\fL=1$. Most
important, whatever it be their form in a given generalized vacuum model,
they must satisfy $\fM(0;r_i)=\fL(0;r_i)=1$ identically, in order to
comply with the cosmic sum rule $\Om^0+\OL^0=1$.

Let us now compute the effective $\wD$ for the generalized vacuum model
introduced above. We proceed as though we would not know that the original
Hubble function is the one given by Eq.\,(\ref{eq:Hgeneral}) and we assume
that it evolves according to the typical expansion rate of the universe
with the DE furnished by a scalar field or some self-conserved entity with
negative pressure:
\begin{equation}\label{H2DE}
H_{\rm D}^2(z)=H_0^2\,\left[\tOm^0\,(1+z)^{\alpha_m}+\,\tOD^0\,\zeta(z)\right]\,,
\end{equation}
where
\begin{equation}\label{eq:defzz}
\zeta(z)\equiv\,\exp\left\{3\,\int_0^z\,dz'
\frac{1+\wD(z')}{1+z'}\right\}\,.
\end{equation}
In Eq.(\ref{H2DE}) we may assume ${\tilde \Omega}_{m}^{0}=0.314 \pm 0.02$
provided by the recent {\it Planck}  results (Ade et al. 2013). The dark
energy density in this picture is given by $\rD(z)=\rD^0\,\zeta(z)$. If
$p_{\rm D}$ denotes the negative pressure associated to the DE, we must
assume (as usual) that $\wD=\rD/p_{\rm D}<-1/3$, in order to grant an
accelerated expansion. The subindex ''D'' in the Hubble function
(\ref{H2DE}) serves to denote the ``DE picture'' description, in contrast
to the original dynamical vacuum energy model, or ``CC picture''
(\ref{eq:Hgeneral}). The tildes in the cosmological parameters in
Eq.\,(\ref{H2DE}) indicate their values in the new picture. Notice that we
also have the corresponding cosmic sum rule $\tOm^0+\tOD^{0}=1$ in the DE
picture, since $\zeta(0)=1$. However, the parameters in the two pictures
need not be identical; in particular, the value of
\begin{equation}\label{eq:DeltaOmega}
\Delta\Om^0\equiv \Om^0-\tOm^0\,,
\end{equation}
even if it is naturally expected small ($|\Delta\Om^0|/\Om^0\ll 1$),  can
play a role in our analysis.

The next point is to implement an important matching condition between the
two expansion histories, namely we require the equality of the expansion
rates of the original dynamical CC picture (\ref{eq:Hgeneral}) and that of
the DE picture (\ref{H2DE}): $H(z)=H_{\rm D}(z)$ ($\forall z$). First of
all we note from equations (\ref{H2DE}) and (\ref{eq:defzz}) that
$\wD(z)=-1+(1/3)\,[({1+z)}/{\zeta}]\,{d\zeta}/{dz}$. From here a
straightforward calculation leads us to a first operative formula to
compute the effective EoS:
\begin{eqnarray}\label{wDE1}
\wD(z)=-1+\frac13\,\alpha_m\,(1+z)^{\alpha_m}\,\,\epsilon(z)\,,
\end{eqnarray}
where
\begin{eqnarray}\label{wDE1epsilon}
\epsilon(z)=\frac{\left[\alpha_m\,(1+z)^{\alpha_m-1}\,\right]^{-1}\,{dE(z)}/{dz}-\tOm^0}{E^2(z)-\tOm^0\,(1+z)^{\alpha_m}}\,.
\end{eqnarray}
It is important to remark that $E(z)$ in the EoS formula must be computed
from (\ref{eq:Hgeneral}), in which the functions $\fM(z;r_i)$ and
$\fL(z;r_i)$ are assumed to be known from the structure of the given
dynamical vacuum model.

It is immediately checked that for the special case of the $\CC$CDM model
($\fM=\fL=1$) and for $\Om^0=\tOm^0$, one has $\epsilon(z)=0$ and
Eq.\,(\ref{wDE1}) reduces to $\wD=-1$, as it should. Therefore, any
departure from this result will be a clear sign that the background
cosmology cannot be one with $\CC=$const. In particular, if the resulting
effective EoS evolves with the expansion, $\wD=\wD(z)$, it will be a sign
of a dynamical vacuum. As we realize now, however -- and this is an
important remark underlying all this work -- the traces of DE dynamics
need not be necessarily attributed to a scalar field with some specific
potential. We will identify later on distinctive features that can emerge
between the dynamics triggered by a running vacuum model and that of a
purely scalar field model. But before unraveling these differences, let us
come back to the above effective EoS formula and show that it can be
brought into a reduced form that is much more convenient for a physical
interpretation.

It turns out that thanks to the constraint imposed by the general Bianchi
identity (\ref{BianchiGeneral}) the EoS formula can be expressed in a more
compact fashion,  directly in terms of the functions $\fM(z)$ and $\fL(z)$
of the generalized vacuum model.  To see this, let us first trade the
cosmic time variable for the redshift variable in
Eq.\,(\ref{BianchiGeneral}), which is easy done using
$d/dt=-(1+z)\,H(z)\,d/dz$, and we find
\begin{equation}\label{BianchiGeneral2}
(1+z)\frac{d}{dz}\,\left[G(z)(\rmr(z)+\rL(z))\right]=\alpha_m\,G(z)\,\rmr(z).
\end{equation}
From this equation and Eq.\,(\ref{eq:Hgeneral}) we arrive at the following
expression:
\begin{eqnarray}\label{eq:Bianchi3}
&&\frac{d}{dz}\,\left[\Om^0\,\fM(z;r_i)(1+z)^{\alpha_m}+\OL^0\,\fL(z;r_i)\right]\nonumber\\
&&=\ \alpha_m\,\Om^0\,\fM(z;r_i)(1+z)^{\alpha_m-1}
\end{eqnarray}
Working out this expression, we find:
\begin{equation}\label{eq:BianchiDiff}
\Om^0\,\fM'(z;r_i)(1+z)^{\alpha_m}+\OL^0\,\fL'(z;r_i)=0\,,
\end{equation}
where the primes denote derivatives with respect to the redshift variable.
The differential relation (\ref{eq:BianchiDiff}) between the functions
$\fM(z)$ and $\fL(z)$ is a reflection of the Bianchi identity, and it
plays a key role to simplify the structure of the effective equation of
state (\ref{wDE1}). Indeed, computing $dE/dz$ from (\ref{eq:Hgeneral}),
and using Eq.\,(\ref{eq:BianchiDiff}), the distinctive dynamical part of
the effective EoS (\ref{wDE1}) can be cast as
\begin{eqnarray}\label{wDE2}
&&\epsilon(z)=\frac{\Om^0\fM(z;r_i)-\tOm^0}{E^2(z)-\tOm^0\,(1+z)^{\alpha_m}}\\
&&=\frac{\Om^0\fM(z;r_i)-\tOm^0}{\left(\Om^0\,\fM(z;r_i)-\tOm^0\right)\,(1+z)^{\alpha_m}+\OL^0\,\fL(z;r_i)}\nonumber
\end{eqnarray}
In this simpler form it becomes transparent that $\wD(z)$ reduces to
$\wD=-1$ for the $\CC$CDM case ($\fM=\fL=1$) provided the parameter
difference (\ref{eq:DeltaOmega}) is zero.

Equation (\ref{wDE2}) leads to an interesting observation. As the function
$\fM$ must satisfy $\fM(0;r_i)=1$, and the parameter difference
(\ref{eq:DeltaOmega}) between the two pictures should be small, it follows
that for any generalized CC model in which $\fM(z;r_i)$ is a monotonous
function of $z$ there will be a point $z=z^{*}$ near our present ($z=0$)
where $\epsilon(z^*)=0$, equivalently $\wD(z^*)=-1$, and hence the
effective EoS function (\ref{wDE1}) will change from $\wD(z)>-1$
(``effective quintessence'') to $\wD(z)<-1$ (``effective phantom'') around
that point, or vice versa. For instance, assume that $\Delta\Om^0<0$ and
that the function $\fM(z;r_i)$ is monotonously increasing with $z$
(therefore decreasing with the expansion). It means that well in our past
$\epsilon(z)>0$, equivalently, $\wD(z)>-1$, so the running vacuum model
behaves there as quintessence. Let us now approach the present time. The
condition $\fM(z;r_i)\to 1$ for $z\to 0$ is satisfied in the manner
$\fM(z;r_i)\gtrsim 1$ , and hence there is a point $z^*$ in our recent
past where the effective EoS entered the phantom regime:
$\epsilon(0\lesssim z\lesssim z^*)<0$. Here we assume that the denominator
of (\ref{wDE2}) stays positive near our present, thanks to the second term
in it, which tends to $\OL^0\fL(0;r_i)=\OL^0>0$ -- in effect larger than
the first term, which is of order $\Delta\Om^0$ near $z=0$ . Similarly, if
$\Delta\Om^0>0$, the phantom regime can only be attained after the present
time, namely when the condition $\fM(z;r_i)\lesssim 1$ is fulfilled (to
the necessary degree) at some $z\lesssim 0$ (in our future). Should,
instead, the function $\fM(z;r_i)$ be monotonically decreasing with $z$
(i.e. increasing with the expansion), then for $\Delta\Om^0<0$ there could
be a crossover from phantom into quintessence in the future, whereas for
$\Delta\Om^0>0$ the same kind of crossover could occur in our recent past.
This is true  only if the denominator of (\ref{wDE2}) stays positive. If
it changes to negative sign in the past but stays positive near our
present, then for  $\Delta\Om^0<0$ we can actually have the opposite
situation, i.e. a transition from quintessence to phantom near our time.
This situation will actually occur in one of our specific examples (cf.
Sect. \ref{sect:numerical} for details).

Needless to say, without giving further details of the parameter values
and of the structure of the generalized vacuum model (\ref{eq:Hgeneral})
-- above all the explicit form of the functions $\fM(z;r_i)$ and
$\fL(z;r_i)$ -- it is not possible to firmly conclude if the
aforementioned crossing will be in our recent past or in our immediate
future. In the next section we shall illustrate various possibilities by
studying a rather general class of running vacuum models, in which the
functions $\fM$ and $\fL$ are given, or can be computed.

\section{Specific running vacuum energy models}\label{sect:runningvac}

We have seen that in the wide class of generalized vacuum models
(\ref{eq:Hgeneral}) we should expect the existence of at least a redshift
point $z=z^{*}$ representing a crossover of the CC divide by the effective
EoS. However, the crossing is not realized by the presence of fundamental
phantom fields in combination with quintessence fields, as amply discussed
within very different points of view in the past (confer e.g. Feng, Wang
\& Zhang 2005, Vikman 2005; Caldwell \& Doran 2005, see also the review by
Copeland, Sami \& Tsujikawa 2006 and references therein). The effective DE
behavior of the EoS is caused, in our case, by the (mild) running of the
vacuum energy density $\rL$ and/or the participation of a (slightly)
anomalous conservation law for matter, sometimes even in combination with
the (slow) running of the gravitational coupling $G$. Such possibility
should be welcome since it shows that the crossover of the CC divide
$\wD=-1$ need not generally be associated to any QFT anomaly, but to the
effective description of what could be a perfectly normal QFT behavior. To
illustrate this fact in concrete terms, in the present section we shall
compute the effective EoS in the context of the three model types
introduced in Sect \ref{sect:dynamicalvac}, all of them being particular
cases of the general structure (\ref{eq:Hgeneral}).

\subsection{Running vacuum $\CC_t$CDM models}\label{sect:RGmatter}

Let us start with the type i) model of Sect. \ref{sect:dynamicalvac}, i.e.
the class  $\CC_t$CDM. We exemplify it with the running vacuum models
whose energy density evolves with the expansion rate in the following way:
\begin{equation}\label{eq:VacuumfunctionCCtCDM}
\rho_{\Lambda}=n_0 +n_2 H^2 +n_{\dot{H}}\,{\dot H}\,.
\end{equation}
Notice that the constant additive term $n_0$ should be the dominant one,
in order that the vacuum energy remains approximately constant near our
present and also during relatively long cosmological intervals of time. In
other words, we expect $n_0\sim \rLo$. We will be more precise later on.
However, the additional $H$-dependent terms endow the vacuum energy with a
dynamical behavior. From the two Friedmann's equations (\ref{friedr}) and
(\ref{friedr2}) it is easy to show that
\begin{equation}\label{eq:ratioH2dotH}
\frac{H^2}{\dot{H}}=-\frac{2}{\alpha_m}\,(1+r)=-\frac{2}{3}\,\frac{1+r}{1+\wm}\,,
\end{equation}
where $r=\rL/\rmr$ is the ratio between vacuum energy density and matter
energy density. This ratio is of ${\cal O}(1)$ at present ($r\sim 7/3$),
whereas in the past $r\to 0$. Therefore, for the relevant epoch
$H^2$ and $\dot{H}$ are dynamical terms of the same order of magnitude
($H_0^2$ is roughly twice $|\dot{H}_0|$). It is only in the remote future
that $r\to\infty$ -- except for the $\CC$XCDM model, where $r$ stays
always bounded (Grande, Sol\`a \& \v{S}tefan\v{c}i\'{c} 2006). As the
dynamical behavior should obviously be mild, we expect that these terms
play a sub-leading role, but at the same time we consider the possibility
that they can have some measurable effect. The coefficients of $H^2$ and
$\dot{H}$ can be parameterized more conveniently in the form
\begin{equation}\label{eq:defn2ndH}
n_2=\frac{3\nu}{8\pi}\,M_P^2\ \ \ \ \ \ \ \ n_{\dot{H}}=\frac{\alpha}{4\pi}\,M_P^2\,,
\end{equation}
$M_P$ being the Planck mass, and where we have traded the dimensionful
parameters $n_2$ and $n_{\dot{H}}$ by the dimensionless ones $\nu$ and
$\alpha$. The theoretical motivation for the structure
(\ref{eq:VacuumfunctionCCtCDM})-(\ref{eq:defn2ndH}) is based on the
renormalization group running. This is also the reason why we have omitted
the linear terms in $H$ (Shapiro \& Sol\`a 2002; Sol\`a 2007; Shapiro \&
Sol\`a 2009). Should the dimensionless parameters $\nu$ and $\alpha$ be of
order one, the vacuum dynamics would be too pronounced and we would have
detected it, so we expect that
\begin{equation}\label{eq:defnualpha}
|\nu|\ll 1\,,\ \   |\alpha|\ll 1\,\,.
\end{equation}
This theoretical expectation is substantiated by particular QFT
calculations (Sol\`a 2007), where $\nu$ is found to satisfy $|\nu|={\cal
O}(10^{-3})$ at most. On the other hand, we can also use
Eq.\,(\ref{eq:VacuumfunctionCCtCDM}) as a phenomenological ansatz and
compare the model with the data so as to extract the maximum allowed
values for these parameters. This has been done in (Basilakos, Polarski \&
Sol\`a, 2012), extending the analysis of (Basilakos et al. 2009; Grande et
al. 2010, 2011; Fabris, Shapiro \& Sol\`a, 2007). These works use the data
on type Ia supernovae (SNIa), the Baryonic Acoustic Oscillations (BAOs),
the shift parameter of the Cosmic Microwave Background (CMB) and the power
matter spectrum. The results suggest that both $|\nu|$ and $|\alpha|$ can
be at most of order $10^{-3}$, in very good agreement with the expectation
(\ref{eq:defnualpha}).
A bound of this order, even if it is relatively tight, could still enable
detection of a mildly time evolving vacuum energy. It is also interesting
that, even with such small values of the parameters, these dynamical
vacuum models could be able to explain the hints presumably detected in
experiments devoted to find evidence of the possible variation of the
so-called fundamental ``constants'' of nature (cf. Fritzsch \& Sol\`a
2012).


The model can be explicitly solved following the procedure outlined in
Sect. \ref{sect:dynamicalvac}. Essentially, one has to solve
Eq.\,(\ref{frie33}) using the explicit form
(\ref{eq:VacuumfunctionCCtCDM}) and the reparameterization
(\ref{eq:defn2ndH}). The computation is straightforward, and the emerging
matter and vacuum energy densities read as follows:
\begin{equation}\label{eq:MatterdensityCCtCDM}
\rho_m(z) =  \rmo~(1 + z)^{3 \xiM}
\end{equation}
and
\begin{equation}\label{eq:CCdensityCCtCDM}
\rL(z)=\rLo+{\rmo}\,\,(\xiM^{-1} - 1) \left[ (1 + z)^{3\xiM} -1  \right]\,,
\end{equation}
with
\begin{equation}\label{eq:defxiM}
\xiM  = \frac{ 1 - \nu }{ 1 - \alpha }\,.
\end{equation}
In the above formula, $\rmo$ is the cold matter density at present.
We have omitted the radiation part, $\rRo$, because it is not
relevant for the study of the EoS behavior near our time. Its
inclusion is nevertheless essential to fit the CMB data. We omit
these details, but for the sake of completeness we quote the
normalized Hubble rate when both the relativistic and
nonrelativistic matter components are included:
\begin{equation}\label{eq:HzCCtCDM}
E^2(z) = \frac{\Omo}{\xiM}~(1+z)^{3 \xiM} +
\frac{\ORo}{\xiR}~(1+z)^{4 \xiR}
                           + \frac{\OLo-\Delta\nu}{1 - \nu}\,,
\end{equation}
where $\Delta\nu=\nu -\bar{\nu}$, with $\bar{\nu}=\alpha\,\Omo+
(4/3)\,\alpha\,\ORo$. Here
\begin{equation}\label{eq:defxiR}
\xiR = \frac{ 1 - \nu }{ 1 - \frac{4}{3}\alpha }\,.
\end{equation}

The current normalized contributions from both relativistic and
nonrelativistic matter, $\Omo$ and $\ORo$, and vacuum energy $\OLo$,
satisfy $\Omo+\ORo+\OLo=1$. The term $\ORo$ can be omitted from that sum
rule when the radiation component is neglected. But the full expression
(\ref{eq:HzCCtCDM}) is indispensable in order to determine the model
parameter values from the CMB data following the procedure of Basilakos et
al. (2012). In this sense, the fit is actually sensitive to two parameters
$(\xiM,\xiR)$, or alternatively $(\nu,\alpha)$. The first set allows a
more compact notation, but the second set is more convenient for the
numerical analysis because any deviation of $\nu$ and/or $\alpha$ from
zero indicates a departure from the $\CC$CDM.

We can indeed check that all the above expressions retrieve their standard
forms when $\nu$ and $\alpha$ are both vanishing (equivalently,
$\xiM=\xiR=1$)\footnote{Notice that in the absence of the condition
$\xiR=1$ the two parameters $\nu$ and $\alpha$ could be large and almost
equal, and this option would still be compatible with $\xiM=1$. However,
the simultaneous constraint $\xiR=1$ is what enforces that $\nu$ and
$\alpha$ must be both small in the physical parameter space.}; in
particular, $\rL$ in (\ref{eq:CCdensityCCtCDM}) then becomes strictly
constant ($\rL=\rLo$) and the cold matter evolution law
(\ref{eq:MatterdensityCCtCDM}) recovers its conventional form
$\rmr(z)=\rmo\,(1+z)^3$, i.e. as in (\ref{solstandardconserv}) for
$\wm=0$.

\begin{figure}
\mbox{\epsfxsize=8.2cm \epsffile{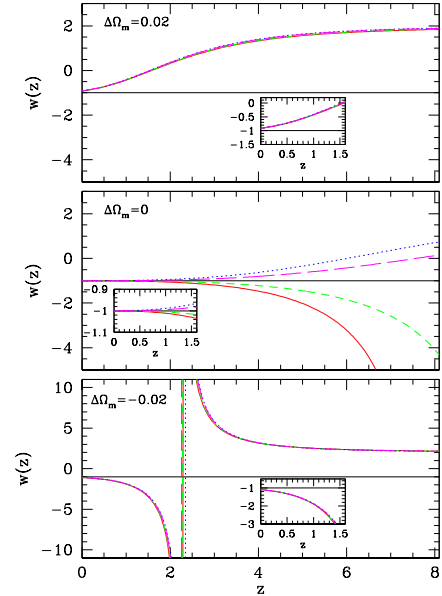}}
\caption{
{\em Outer Plots:} The $\CC_t$CDM evolution of the effective EoS parameter
for various $\Delta \Omega_{m}^{0}$ values. We assume $\alpha= 3\nu/4$ (see Sect. 5 for details).
To produce the lines we used
$\nu=-0.002$ (blue/dotted), $\nu=-0.001$ (magenta/long dashed),
$\nu=0.001$ (green/dashed) and $\nu=0.002$ (red/solid).
{\em Inner Plots:}
detail of the evolution of $w_{D}(z)$ in the shorter range  $0\leqslant z\lesssim 1.5$ relevant for SNIa observations.}
\end{figure}

Inserting equations (\ref{eq:MatterdensityCCtCDM}) and
(\ref{eq:CCdensityCCtCDM}) in Eq.\,(\ref{eq:Hgeneral}), and comparing with
Eq.\,(\ref{eq:Hgeneral2}) in the nonrelativistic matter epoch ($\wm=0$),
we can explicitly identify the functions $\fM(z;r_i)$ and $\fL(z;r_i)$ for
the dynamical vacuum model (\ref{eq:VacuumfunctionCCtCDM}). They read
\begin{eqnarray}\label{eq:fMfLCCtCDM}
\fM(z)&=&  (1+z)^{3(\xiM-1)}=(1+z)^{3\frac{\alpha-\nu}{1-\alpha}} \\
\fL(z) &=& 1+\frac{\Omo}{\OLo}\,(\xiM^{-1} - 1)\left[ (1 + z)^{3\xiM} -1  \right]\nonumber\\
&=& 1+\frac{\Omo}{\OLo}\,\frac{\nu-\alpha}{1-\nu}\left[ (1 +
z)^{3\frac{1-\nu}{1-\alpha}} -1  \right]\label{eq:fMfLCCtCDM2}\,.
\end{eqnarray}
As expected, they satisfy $\fM(0)=1=\fL(0)$ at $z=0$. Notice that that
these functions are of the form $\fM(z;\alpha,\nu)$ and
$\fL(z;\alpha,\nu)$, i.e. they depend on two independent parameters
$\alpha$ and $\nu$. It is interesting to note that despite they are
expected to be small in absolute value, the sign difference $\alpha-\nu>0$
or $\alpha-\nu<0$ determines whether $\fM(z)$ increases or decreases with
$z$, and this feature may determine the effective quintessence or phantom
behavior of the model (cf. Sect. \ref{sect:effectiveEOS}). From
Eq.\,(\ref{eq:BianchiDiff}) it is obvious that when $\fM(z;r_i)$
increases, $\fL(z;r_i)$ must decrease, and vice versa. We can check that
this is indeed the case for the specific functions of the model under
consideration. Finally, substituting the expressions (\ref{eq:fMfLCCtCDM})
and (\ref{eq:fMfLCCtCDM2}) in Eq.\,(\ref{wDE2}) and rearranging, we find
$$
\epsilon(z)=\frac{\xi\left[\Omo(1+z)^{3(\xiM-1)}-\tOm^0\right]}{\,\Omo\left[(1+z)^{3\xiM}-1\right]-\xi\,\left[\tOm^0(1+z)^{3}-1\right]}\,,
$$
\begin{equation}\label{eq:wDECCtCDM}
\phantom{XXXXXXXXXXXXXXXXXXXX}
\end{equation}
where we recall that $\xiM$ can be written in terms of $(\nu,\alpha)$
through Eq.\,(\ref{eq:defxiM}). We note immediately that $\epsilon(z)=0$
($\forall z$) if $\nu=\alpha=0$ and if the parameter difference
(\ref{eq:DeltaOmega}) vanishes, as expected. Moreover, if
$\Delta\Omega^0_m=0$, we have $\epsilon(0)=0$, equivalently $\wD(0)=-1$,
for any value of $\nu$ and $\alpha$. This does not apply, of course, for
$z\neq 0$. It is instructive to expand $\epsilon(z)$ linearly in the small
parameters $\nu$ and $\alpha$, assuming that $z$ is not very large (i.e.
for points relatively close to our current universe) and with the natural
assumption $|{\Delta\Omo}/{\Omo}|\ll1$. After substituting the result in
Eq.\,(\ref{wDE1}), we arrive at the following approximate effective EoS
$$\wD(z)\simeq
-1+\frac{\Omo}{\OLo}(1+z)^3\left[\frac{\Delta\Omo}{\Omo}+3(\alpha-\nu)\,\ln(1+z)\right]\,.
$$
\begin{equation}\label{eq:EoSCCtCMDparamsmall}
\phantom{XXXXXXXXXXXXXXXXXXXX}
\end{equation}
This equation, even though only approximate, reveals the essential
qualitative facts of the effective EoS for the present model. For example,
we confirm that we can recover the $\CC$CDM limit, $\wD(z)=-1$, for
$\nu=\alpha=0$ and $\Delta\Omo=0$, as it should be; and that $\wD(0)=-1$
irrespective of $\nu$ and $\alpha$. Moreover, for $\Delta\Omo=0$ and
nonvanishing $\alpha$ and $\nu$, the effective EoS mimics quintessence
($\wD\gtrsim -1$) for $\alpha-\nu>0$, and phantom DE ($\wD\lesssim-1$) for
$\alpha-\nu<0$. If, however, there is a significant relative deviation
between the parameters $\Omo$ and $\tOm^0$, the situation could change,
depending on the size and sign of the term ${\Delta\Omo}/{\Omo}$. It is
easy to check that with the current values of the cosmological parameters
and their precisions we can easily get $\wD=-1\pm0.1$ for supernovae data
at $z\simeq 1$ (cf. Fig. 1 and Sect.\ref{sect:numerical} for the detailed
numerical analysis). Therefore, both the quintessence and phantom regime
can be accounted for in an effective way by the dynamical class of
$\CC_t$CDM models, without invoking fundamental scalar or phantom fields
as responsible for the DE. Interestingly, the effective EoS of the
dynamical vacuum model does not, though, adapt to the simple
parameterizations of the DE in vogue in the literature, which cannot
describe this kind of scenarios. We shall come back to this issue in
Sect.\ref{sect:simpleEoS}. {Let us also note that the kind of
effective EoS plots obtained here (and the impact on them from the
assumptions of matter density) are also observed in generic approaches
where one reconstructs the equation of state of dark energy (see e.g.
Sahni, Shafieloo, \& Starobinsky, 2008)}.

\begin{figure}
\mbox{\epsfxsize=8.2cm \epsffile{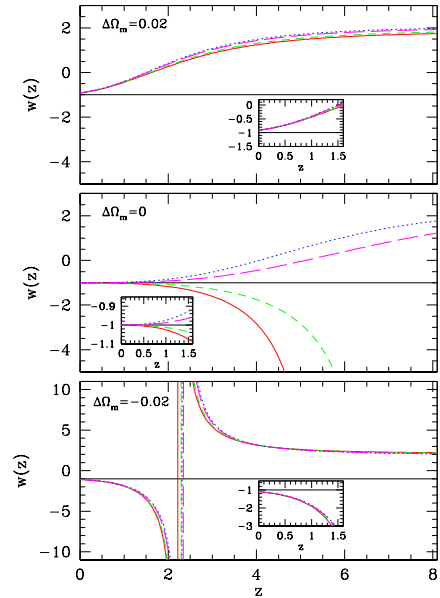}}
\caption{
The $\CC_t$$G_t$CDM evolution of the effective EoS parameter. For more
details concerning the curves see caption of Fig. 1}
\end{figure}

\subsection{$\CC_t$$G_t$CDM models: running $\CC$ and $G$}\label{sect:runnCCtGt}

We discuss now, more briefly, the type ii) model introduced in Sect.
\ref{sect:dynamicalvac}, i.e. the  $\CC_tG_t$CDM model. In this case,
matter is covariantly conserved and the Bianchi identity can be fulfilled
through a dynamical interplay between a running vacuum energy and a
running gravitational coupling, see Eq.\,(\ref{Bianchi1}). Unfortunately,
in general in this class of models it is not possible to provide a simple
analytical solution of the cosmological equations, especially if we start
again with a vacuum dynamical law of the full form as indicated in
Eq.\,(\ref{eq:VacuumfunctionCCtCDM}), i.e. containing both $H^2$ and
$\dot{H}$ terms. We have, however, seen from the previous section that the
two dynamical terms, $H^2$ and $\dot{H}$, play a similar role in the
effective EoS.  We will assume that this is also the case here. Moreover,
as we discussed in Eq.\,(\ref{eq:ratioH2dotH}), $H^2$ and $\dot{H}$ are of
the same order of magnitude at present and in the past. Therefore it
should suffice to focus here on the model type
(\ref{eq:VacuumfunctionCCtCDM}) under the assumption that $n_{\dot{H}}=0$
-- equivalently, $\alpha=0$ (as in Shapiro, Sol\`a \&
\v{S}tefan\v{c}i\'{c}, 2005; Grande et al. 2010; Grande et al. 2011). Then
we are left with a single parameter, $n_2$ (or $\nu$) and this simplifies
the analysis and the confrontation of the model with the data. Such
assumption enables also to obtain a relatively simple analytical
expression for the effective EoS of this model, as we shall see below.

For the $\CC_tG_t$CDM model, the functions $\fM$ and $\fL$ involved
in the expansion rate (\ref{eq:Hgeneral2}) can be written as
follows:
\begin{equation}\label{eq:functionsfMfLCCtGtCDM}
\fM(z)=\frac{G(z)}{G_0}\equiv g(z)\,,\ \ \ \ \
\fL(z)=\fM(z)\,\frac{\OL(z)}{\OLo}\,,
\end{equation}
where $G_0\equiv 1/M_P^2$ is the current value of the gravitational
coupling, and $\OL(z)=\rL(z)/\rco$ is the vacuum energy density normalized
to the current critical density. Clearly, $\fM(0)=1=\fL(0)$ is fulfilled
as a basic normalization condition that these functions were supposed to
satisfy. The function $g(z)$ is the dimensionless gravitational coupling
at any cosmic epoch, normalized to the current Newton's constant. Solving
Eq.\,(\ref{Bianchi1}) in combination with (\ref{friedr}) it can be exactly
determined as a function of the
 normalized Hubble rate $E(z)=H(z)/H_0$:
\begin{equation}\label{GH}
g(E)=\frac{1}{1+\nu\,\ln{E^2}}\,.
\end{equation}
The logarithmic behavior clearly
shows that $G$ varies very slowly with the cosmic
evolution. If we focus on the nonrelativistic epoch, where the EoS is
measured, the corresponding solution in terms of the redshift can be
obtained in the form of an implicit equation:
\begin{eqnarray}\label{eq:fMCCtGtCDM}
\frac{1}{g(z)}-1+\nu\,\ln\left[\frac{1}{g(z)}-\nu\right]
=\nu\,\ln\left[\Om(z)+\OLo-\nu\right]\,.
\end{eqnarray}
Similarly an expression for $\fL(z)$, which hinges on the previous one,
ensues:
\begin{eqnarray}\label{eq:fLCCtGtCDM}
\fL(z)=\frac{g(z)}{1-\nu\,g(z)}\,\left\{1+\frac{\nu}{\OLo}\,\left[\Om(z)\,g(z)-1\right]\right\}\,.
\end{eqnarray}
Here $\Om(z)=\rmr(z)/\rco=\Omo\,(1+z)^3$ is the matter energy density
normalized to the current critical density. Notice from
(\ref{eq:fMCCtGtCDM}) and (\ref{eq:fLCCtGtCDM}) that we get $g=1$ and
$f=1$ identically for $\nu=0$. Moreover, for any $\nu$ we can easily check
the correct normalizations $\fM(0)=g(0)=1$ and $\fL(0)=1$ in these
explicit formulas, after recalling that $\Omo+\OLo=1$.

Finally, we can insert the above expressions in the general
effective EoS equation (\ref{wDE1}),(\ref{wDE2}) so as to obtain the
desired result for the $\CC_t$$G_t$CDM model:
$$
\epsilon(z)=\frac{\left(1-\nu g(z)\right)\left(\Omo
g(z)-\tOm^0\right)}{\left[\Omo g(z) - \tOm^0+\nu
g(z)\tOm^0\right](1+z)^3 + g(z) \left(\OLo-\nu\right)}\,.
$$
\begin{equation}\label{eq:wDECCtGtCDM}
\phantom{XXXXXXXXXXXXXXXXXXXX}
\end{equation}
Note that in order to evaluate this expression exactly one has to obtain
first $g(z)$ by numerically solving the implicit equation
(\ref{eq:fMCCtGtCDM}). It may, however, be illuminating to expand
equations (\ref{eq:fMCCtGtCDM}) and (\ref{eq:wDECCtGtCDM}) linearly in the
small parameter $\nu$, assuming that $z$ is not very large (i.e., once
more for points relatively close to our current universe) and for
$|{\Delta\Omo}/{\Omo}|\ll1$. The final result for the effective EoS, in
this approximation, is the following:
$$\wD(z)\simeq
-1+\frac{\Omo}{\OLo}(1+z)^3\left[\frac{\Delta\Omo}{\Omo}-\nu\,\ln\left[\Om(z)+\OLo\right]\right]\,.
$$
\begin{equation}\label{eq:EoSCCtGtCMDparamsmall}
\phantom{XXXXXXXXXXXXXXXXXXXX}
\end{equation}
Once more, using the present data, we can easily recreate scenarios
leading to $\wD=-1\pm0.1$, thus mimicking both the quintessence and
phantom regimes in a purely artificial way, namely without calling upon
fundamental quintessence or phantom fields (cf. Fig. 2 and
Sect.\ref{sect:numerical} for the detailed numerical analysis).

\subsection{Class $\CC$$G_t$CDM: running $G$ with anomalous matter conservation law}\label{sect:runnGt}

Our final example is the type iii) class of models of Sect.
\ref{sect:dynamicalvac}, or $\CC G_t$CDM models. Here the original vacuum
term is not evolving with the expansion, but the effective description of
this scenario leads to a case of virtually dynamical DE. The effect is
caused by the fact that matter is not conserved in this kind of model, and
this can be covariantly consistent provided there is a small running of
the gravitational coupling. As a result, two anomalous functions $\fM\neq
1$ and $\fL\neq 1$ will be generated in Eq.\,(\ref{eq:Hgeneral}), implying
a nontrivial evolution of the effective DE. To illustrate the present
class, we take a model of this sort that was discussed in (Fritzsch \&
Sol\`a, 2012) within the framework of dynamical vacuum energy, and in
(Guberina, Horvat \& Nikolic, 2006) in the holographic context. The model
assumes an anomalous evolution law for matter, of the form
\begin{equation}\label{eq:MatterdensityCCGtM}
\rho_m(z) =  \rmo~(1 + z)^{3 (1-\eta)}\,,
\end{equation}
which is similar to Eq.(\ref{eq:MatterdensityCCtCDM}), but now with no
accompanying variation of the vacuum energy.

\begin{figure}
\mbox{\epsfxsize=8.2cm \epsffile{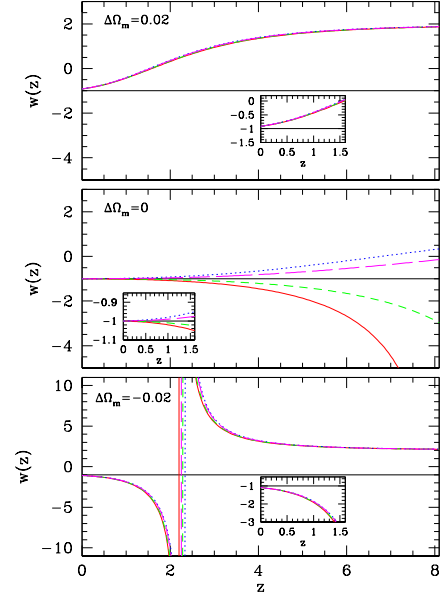}}
\caption{
The $\CC$$G_t$CDM evolution of the effective EoS parameter. Remaining details as in Fig. 1.}
\end{figure}

The coefficient $\eta$ that appears in Eq.(\ref{eq:MatterdensityCCtCDM})
has a different interpretation to that of $\nu$ for the $\CC_t$CDM and
$\Lambda_{t} G_{t}$CDM models of sections \ref{sect:RGmatter} and
\ref{sect:runnCCtGt} since the CC remains constant here. Even so, for
convenience we adopt the same notation $\nu \equiv \eta$.
The anomalous law (\ref{eq:MatterdensityCCGtM}) for matter conservation
was first considered in (Shapiro \& Sol\`a, 2003) and then also in (Wang
\& Meng, 2005) and (Alcaniz \& Lima, 2005).

Next we compute the effective EoS of this model. The corresponding
function $\fM$ reads as follows:
\begin{equation}\label{eq:functionsfMfLCCGtCDM1}
\fM(z)=\fL(z)\,(1+z)^{-3\nu}\,,
\end{equation}
where $\fL(z)$ in this formula is given by
\begin{equation}\label{eq:functionsfMfLCCGtCDM2}
\fL(z)=\frac{G(z)}{G_0}=\,\left[\Omo\left(1+z\right)^{3(1-\nu)}+\OLo\right]^{\nu/(1-\nu)}\,.
\end{equation}
The explicit form for $G(z)$ in the last expression follows from solving
the differential equation (\ref{dGneqo}) in the matter epoch. With the
help of (\ref{wDE2}) and the previous expressions the effective EoS of
this model becomes completely determined. Its dynamical part reads
\begin{eqnarray}\label{eq:EoSCCGtCDM}
\epsilon(z)=\frac{\Om^0\fL(z)\,(1+z)^{-3\nu}-\tOm^0}{\left[\Om^0\fL(z)\,(1+z)^{-3\nu}-\tOm^0\right]\,(1+z)^{3}+\OL^0\,\fL(z)}\,.\nonumber
\end{eqnarray}
\begin{equation}\label{eq:wDECGGtCDM}
\phantom{XXXXXXXXXXXXXXXXXXXX}
\end{equation}
Expanding this formula linearly in the small parameter $\nu$ for not very
high $z$ and $|{\Delta\Omo}/{\Omo}|\ll1$, the effective EoS can be
approximated as follows:
$$\wD(z)=-1+\frac{\Omo}{\OLo}\,\,(1+z)^{3}\left\{\frac{\Delta\Omo}{\Omo}+\nu\,\ln\left[\Omo+\frac{\OLo}{(1+z)^3}\right]\right\}\,.
$$
\begin{equation}\label{eq:EoSCCGtCDMapprox}
\phantom{XXXXXXXXXXXXXXXXXXXX}
\end{equation}
For a numerical example, see Fig. 3. In Sect.\ref{sect:numerical} we
present the details of the numerical analysis.

\subsection{Some usual parameterizations of the effective EoS}\label{sect:simpleEoS}

We may now briefly compare the various forms for the effective EoS
obtained in the previous sections with the most usual ones employed in the
literature and in the practical analysis of the cosmological data, say
from WMAP (Komatsu et al. 2011) and {\it Planck}  (Ade et al. 2013).

Popular parameterizations have been proposed and tested in the past in the
literature. They intended to be simple and as useful as possible for a
large class of models, but they have their own limitations. As an example,
consider the
very simple form
\begin{equation}\label{weparam1}
\wD(z)=\omega_0+\omega_1 z\,. \
\end{equation}
This form is conceivable only for recent data since its linear behavior
with the redshift cannot be extended to the early universe. A
parametrization that overcomes this difficulty is the CPL parametrization
(Chevallier \& Polarski 2001; Linder 2003):
\begin{equation}\label{weparam2}
\wD(a)=\omega_0+\omega'_0 (1-a)\,,
\end{equation}
where $a$ is the scale factor. Finally, let us
quote \,(see Jassal et al. 2005; 2010)
\begin{equation}\label{weparam3}
\wD(z)=\omega_0+\omega'_0\,\frac{z}{(1+z)^p}\,, \ \ \ (p=1,2,...)\,,
\end{equation}
where $\omega'_0=(d\wD/dz)_{z=0}$ is assumed to be significantly smaller
than $\omega_0$ (in absolute value), i.e. $|\omega'_0/\omega_0|\ll 1$, as
$\omega'_0$ controls the dynamical part of the DE. One naturally expects
that $\omega_0$ is close to $-1$, i.e. $\omega_0=-1+\delta$, with
$|\delta|\ll 1$. It is clear that (\ref{weparam3}) is a generalization of
(\ref{weparam2}) since for $p=1$ they coincide. Notice that the asymptotic
behavior of $\wD(z)$ in this kind of parametrizations is smooth: for $p=1$
one has $\wD(+\infty)=\omega_0+\omega'_0$, and for $p\geq2$,
$\wD(+\infty)=\omega_0$. Let us consider the corresponding Hubble rate.
For example, for $p=1$ one obtains from
equations\,(\ref{H2DE})-(\ref{eq:defzz}) the following result:
\begin{eqnarray} \label{Hzzzquint}
E^2(z)&=&
\tOm^0\,(1+z)^{3}\\
&+&\tOD^0\,(1+z)^{3(1+\omega_0+\omega'_0)}\,\exp\left[-3\,\omega'_0\,\frac{z}{1+z}\right]\nonumber \,.
\end{eqnarray}
Even though the parameter family of EoS (\ref{weparam3}) is quite general
and useful for many models, the kind of running vacuum models under
consideration cannot be described by it. The expansion history traced by
the Hubble function (\ref{Hzzzquint}) is still too simple to cover the
behavior of the dynamical vacuum models discussed in the previous
sections. Both at small and at high redshift redshift the DE part of
(\ref{Hzzzquint}) basically behaves as a power $\sim
(1+z)^{3(1+\omega_0+\omega'_0)}$, with just a small correction from the
exponential since $|\omega'_0/\omega_0|\ll 1$. In other words, it roughly
takes on the form $\sim(1+z)^{3(\delta+\omega'_0)}$. This expression can
effectively behave as a CC term (as suggested by observations), provided
$|\delta|$ and $|\omega'_0|$ are small, as indeed assumed. This is at
least the logic followed by the usual parameterizations of the DE. For
example, for $p=1$ Eq.\,(\ref{weparam3}) exhibits the approximate behavior
\begin{equation}\label{weparam3b}
\wD(z)\simeq -1+\delta+\omega'_0\,\frac{z}{1+z}\,.
\end{equation}
The latter describes quintessence  for $\delta+{\omega'_0\,z}/(1+z)>0$ and
phantom energy for $\delta+{\omega'_0\,z}/(1+z)<0$. While this is of
course only a parametrization, at the fundamental level one usually
attributes this behavior to some dynamical scalar field, which is supposed
to underlie the EOS (\ref{weparam3}). In fact, it is well-known (cf.
Peebles \& Ratra 2003) that a dynamical scalar field yields a contribution
to the expansion rate of the form given in Eq.\,(\ref{H2DE}). In point of
fact, within the quintessence-like scenarios one assumes that the DE
behavior is fully generated in that way (i.e. neglecting the existence of
the CC term \textit{ab initio}).

However, as we have seen in the various examples provided in the previous
section, we can actually reproduce all kinds of quintessence or
phantom-like behaviors from a dynamical model in which the vacuum density
and/or the gravitational constant are running. In such framework, which is
conceptually quite different from the one usually adopted in the
literature, the fundamental scalar fields are not primarily responsible
for the DE, and in particular there is no need of fundamental phantom
fields (i.e. fields with negative kinetic term). For example, the
dynamical vacuum models that could be ultimately responsible for the
generic behavior (\ref{eq:VacuumfunctionCCtCDM}) might emerge from the
expected quantum effects of the effective action of QFT in curved
spacetime (Sol\`a, 2013). Specifically, in an expanding universe these
quantum effects should endow the vacuum energy density of a time evolution
inherited from additional (even) powers of the Hubble rate (Shapiro \&
Sol\`a 2002 and 2003; Sol\`a 2007; Shapiro \& Sol\`a 2009). In this kind
of framework the scalar (and fermion) fields can just enter through their
virtual quantum effects that renormalize the vacuum energy density, all of
them with standard kinetic terms.

Furthermore, it is pretty obvious that the EoS behaviors inferred for the
various dynamical vacuum models considered in the previous sections,
cannot be accommodated into the relatively simple form (\ref{weparam3b})
or to any of the initial expressions (\ref{weparam1}), (\ref{weparam2}) or
(\ref{weparam3}). This is particularly transparent when we compare the
approximate forms (\ref{eq:EoSCCtCMDparamsmall}),
(\ref{eq:EoSCCtGtCMDparamsmall}) and (\ref{eq:EoSCCGtCDMapprox}) with
(\ref{weparam2}) or (\ref{weparam3b}). But these are nevertheless the ones
used e.g. for the WMAP and {\it Planck}  analysis of the cosmological
parameters.

\section{Numerical analysis of the effective equation of state} \label{sect:numerical}

In the following we show the evolution of the effective EoS parameter for
the three vacuum models considered in the text, $\CC_t$CDM (Fig. \,1),
$\CC_t$$G_t$CDM (Fig.\,2) and $\CC$$G_t$CDM (Fig.\,3). Note that we sample
$\Delta \Omega_{m}^{0} \in [-0.02,0.02]$ in steps of 0.02 that corresponds
to the $1\sigma$ {\it Planck}  error (Ade et al. 2013). We remind the
reader that for the DE models we use ${\tilde \Omega}_{m}^{0}=0.314$ (Ade
et al. 2013) which implies that the corresponding $\Omega_{m}^{0}$ of the
vacuum models lies in the interval $[0.294,0.334]$. Owing to the fact that
$|\nu|$ is found to be rather small when the vacuum models are confronted
to the cosmological data, $|\nu|\leq{\cal O}(10^{-3})$ (Basilakos et al.
2009; Grande et al. 2011; Basilakos et al. 2012), we decide to use four
different values of $\nu$ around $\nu=0$ -- the latter providing the
$\CC$CDM result. As an example, the lines in the figures correspond to
$\nu=-0.002$ (blue/dotted), $\nu=-0.001$ (magenta/ long dashed),
$\nu=0.001$ (green/dashed) and $\nu=0.002$ (red/solid).

Generally, as it can be seen for $\Delta \Omega_{m}^{0} \ne 0$ (see upper
and lower panels in Figs. 1, 2 and 3) the effective EoS parameter
$w_{D}(z)$ shows almost the same behavior for all three vacuum models.
This is not the case for the $\CC_t$CDM model with $\Delta \Omega_{m}^{0}
=0$. In particular the main comparison results can be summarized in the
following statements (for nomenclature of models, see sections 2 and 4):
\begin{itemize}

\item {\bf $\CC_t$CDM model}  As we said, this model has, in
    principle, two independent parameters, $\nu$ and $\alpha$.
    However, one can actually show that BBN bounds imply
    $|\nu-4\alpha/3|< {\cal O}\left(10^{-3}\right)$ (Sol\`a, 2012).
    The simplest possibility is to take $\alpha= 3\nu/4$, and this is
    what we will do for  most of the numerical analysis, except for an
    special case considered at the end of this section. Notice that
    $\alpha= 3\nu/4$ is tantamount to take a strictly standard
    behavior for the radiation density, $\rho_r\sim a^{-4}$ (as
    assumed in Basilakos, Polarski \& Sol\`a, 2012).

We find the following situations (see Fig. 1):

    a) For $\Delta \Omega_{m}^{0} =0.02$, the effective EoS parameter
    remains always $w_{D}(z) > -1$ as well as it tends to
    $w_{D}(0)\simeq -0.9$ at the present time (see the detail in the
    magnified part of Fig. 1 in the interval $0\leqslant z\lesssim
    1.5$). It is therefore a quintessence-like behavior in that
    region. We also find that for $z\ge 1.5$ the EoS parameter is
    positive, which means that the cosmic expansion in this case is
    more rapidly decelerated than in the usual $\Lambda$CDM
    cosmological model. As a result the acceleration process is a bit
    retarded as compared to the standard model. {Notice that in the
    far future the EoS parameter tends to $-1$, as seen from
    Eq.\,(\ref{wDE1}) and (\ref{eq:wDECCtCDM});}

    b) In the special situation where $\Delta \Omega_{m}^{0}$ is {
    strictly equal to $0$, the effective EoS remains always either
    phantom (for $\nu>0$) or quintessence (for $\nu<0$) for $z\ge 0$.
    In all these cases the current value is exactly $w_{D}(0)= -1$. It
    interesting to mention    that for $\nu>0$ we find a transition
    from phantom into quintessence in the far future ($z\simeq
    -0.5$);}

    c) In the case $\Delta \Omega_{m}^{0} =-0.02$, the evolution of
    the EoS parameter is in the phantom regime with $w_{D}(0)\simeq
    -1.09$, i.e. slightly below $-1$ for $\nu<0$. Before reaching
    $z=0.5$ it already takes the value $w_{D}\lesssim -1.2$. These are
    precisely the kind of phantom behaviors we usually encounter in
    the analyses of WMAP (Spergel et al. 2007; Komatsu et al. 2011)
    and {\it Planck}  (Ade et al. 2013) data.
    On the other hand, close to $z\simeq 2$ one can see a kind of
    divergent feature due to the fact that the denominator of
    Eq.(\ref{eq:wDECCtCDM}) vanishes at this point. We would like to
    stress that there is no real singularity at this point because the
    assumed fundamental RG model behaves smoothly for all values of
    redshifts, Eq.(\ref{eq:Hgeneral}), i.e. all densities are
    perfectly finite at this point. It is only the effective EoS
    description of the original RG model that points to this fake
    singularity, which is nothing but an artifact of the
    parametrization of the running vacuum model as if it were a DE
    model with expansion rate (\ref{H2DE})\,\footnote{The presence of
    this kind of fake singularities has also been observed in the
    effective EoS description of other DE models in the literature,
    see e.g. Shafieloo \textit{et al.} 2005; Sahni \& Shtanov 2005 and
    2002, as well as in Sol\`a \& \v{S}tefan\v{c}i\'{c} 2005 and
    2006.}. From the observational point of view, it is interesting to
    mention that if we would identify a sort of anomaly like the above
    when comparing with the the future $w(z)$ data form the the next
    generation of surveys (based on {\em Euclid} satellite, Cimatti
    \textit{et al.}, Laureijs \textit{et al.} 2011), then we could
    suspect that there is no fundamental dynamical field behind the
    EoS but something else, in particular the RG model under
    consideration. Finally, at very large redshifts $z\gg 10$, we find
    $w_{D}(z)\to 0$. {Once more we find $w_{D}(z\to -1) \rightarrow
    -1$.} Next we describe briefly the results found for the other two
    models, the details are in Figs. 2 and 3.

\item  {\bf $\CC_t$$G_t$CDM model} (cf. Fig. 2): in this case
{we observe that the behavior of the $\CC_t$$G_t$CDM effective EoS
parameter is similar to that of $\CC_t$CDM [for comparison see cases
(a), (b) and (c) in Fig.1].}
Once more, for $\Delta \Omega_{m}^{0} =0$,  the model behaves as
phantom (for $\nu<0$), and as quintessence (for $\nu>0$). The
evolution into these regimes with increasing $z$ is actually faster
than in the $\CC_t$CDM  case studied before.

\item  {\bf $\CC$$G_t$CDM model} (see Fig. 3): this case behaves
    qualitatively more similarly to the $\CC_t$GCDM vacuum model,
    although the quantitative differences are manifest, especially for
    $z \ge 1$. Both of these models share the non-conservation of
    matter, which is compensated by the running of $G$ and $\CC$
    respectively.

\end{itemize}

\begin{figure}
\mbox{\epsfxsize=8.2cm \epsffile{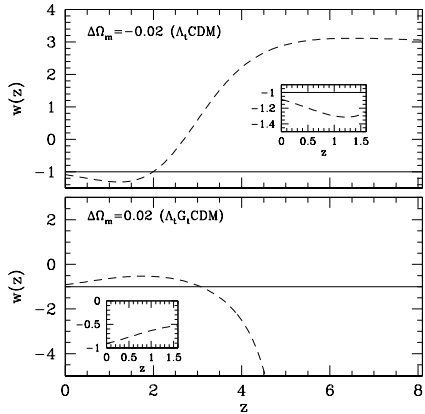}}
\caption{
{\em Upper Panel:}The $\CC_t$CDM evolution of the effective EoS parameter
for $\Delta \Omega_{m}^{0}=\nu=-0.02$.
{\em Lower Panel:} $w_{D}$ versus $z$ for
$\CC_t$$G_t$CDM $\Delta \Omega_{m}^{0}=\nu=0.02$.
{\em Inner Plots:}
The corresponding $w_{D}(z)$ for $z\lesssim 1.5$.}
\end{figure}

Before closing the numerical analysis of the various models, let us
address a couple of special situations. First, assume relatively large
values of  $|\nu|={\cal O}(10^{-2})$. These are not favored by the
cosmological data within the simple class of running models examined here
(Basilakos et al. 2009; Grande et al. 2011; Basilakos et al. 2012), but
could be accommodated within some generalized versions such as the
$\CC$XCDM models (Grande, Sol\`a \& \v{S}tefan\v{c}i\'{c}, 2006, 2007).
Let us analyze this case, as it will serve to illustrate the possibilities
at our disposal. We find that the general $w_{D}(z)$ evolution for the
majority of the vacuum models is similar to those of $|\nu|={\cal
O}(10^{-3})$ with only two exceptions. Specifically, the first model is
the $\CC_t$CDM with $(\Delta \Omega_{m}^{0},\nu)=(-0.02,-0.02)$ (upper
panel of Fig.4). The passing to the $w_{D}(z)<-1$ phase took place at
$z\simeq 2$ where the $\CC_t$CDM model switched from the quintessence- to
the phantom-like EoS. {Interestingly, at $z\simeq 0.7$ which is the
redshift that the universe enters the accelerated period, the effective
EoS takes the value $w_{D}(0.7)\simeq -1.25$}. The current value of the
effective EoS parameter behavior is $w_{D}(0)\simeq -1.09$. The second
vacuum model is the $\CC_t$$G_t$CDM with $\Omega_{m}^{0}=\nu=0.02$ (lower
panel of Fig.4). In this case we meet an alternative crossover from
phantom into quintessence at $z\simeq 3$, sustained until the present
epoch, where $w_{D}(0)\simeq -0.9$. The effective quintessence phase
therefore spans the entire interval where low and high redshift supernovae
have been measured.

{The second special case we would like to consider is when we keep the
$\CC_t$CDM parameters $\nu$ and $\alpha$} within the strictly allowed
range for these models, namely $|\nu, \alpha|={\cal O}(10^{-3})$, but we
assume that the future accuracy of the measurements has reached
$|\Delta\Omega_m^0|=0.005$ (rather than $0.02$ as it is right now). Notice
that we can relax here the assumption $\alpha=3\nu/4$ provided $|\nu|$ and
$|\alpha|$ are both small as indicated above. Two specific examples of
this situation are depicted in Fig. 5.  The interesting feature here is
that we have an effective transition from quintessence into phantom regime
near our time, namely at around $z=1.5$ (the larger is $\alpha>0$ the
closer is the transition to $z=0$), and therefore accessible to future DE
surveys such as the WFIRST project (Green {\textit{et al.} 2011; D.
Spergel {\textit{et al.} 2013). A similar dynamical situation occurs for
the $\CC_t$$G_t$CDM model, although in this case the transition from
quintessence to phantom is significantly earlier in cosmic time (cf. lower
panel of Fig.5). On the other hand, if $\Delta\Omega_m^0=0.005$ then the
corresponding effective EoS parameter of the different vacuum models is
approximated by the upper panels of Figs.1-3.

Finally, we would like to stress that in order to investigate whether the
expansion of the observed universe follows one of the above possibilities,
we need a robust extragalactic equation of state indicator at redshifts
$z\ge 1$. Such high quality $w(z)$ data are expected to be available from
the future surveys like the aforementioned WFIRST project and the {\em
Euclid} mission (Cimatti \textit{et al.}, Laureijs \textit{et al.}  2011).

\section{Discussion and conclusions}

In this work we have shown that there is a wide class of models with
variable vacuum energy density $\rL$ and/or gravitational coupling $G$
that can mimic the behavior of quintessence and phantom fields. These
dynamical vacuum models generally lead to a non-trivial effective equation
of state (EoS) which cannot be described by the usual parameterizations.
The usual observational procedures to pin down the empirical form of the
EoS for the DE in the current generation of precision cosmology
experiments should keep in mind this possibility. While the precise
dynamics behind $\rL$ is not known, various works in the literature
suggest that a fundamental $\rL$ can display a running in QFT in curved
spacetime which can be translated into redshift evolution. This evolution
endows the effective EoS of a behavior which can help explaining the
observation of possible ``mirage transitions'' from quintessence to
phantom DE, or vice versa, without relying at all on the existence of
fundamental quintessence or phantom scalars fields.

{In this work we presented some theoretical possibilities that can
fit the cosmological data pretty well but that are not covered by popular
EoS dark energy parametric forms such as the well known CPL one
((Chevallier \& Polarski 2001; Linder 2003)). This is important since we
do not know what is the actual model of DE and using an inappropriate
parametric form of it to fit the data can become misleading. We have also
shown that assuming different values of matter density can substantially
affect the form of the effective $w(z)$. While the concept of cosmographic
degeneracy has been discussed for a long time in the literature, where it
is shown how assumptions of matter density or curvature can affect the
reconstructed equation of state of dark energy (Clarkson, Cortes, \&
Bassett, 2007; Sahni, Shafieloo, \& Starobinsky, 2008; Shafieloo \&
Linder, 2011).} still it is important to highlight this issue that working
with the EoS of DE can be very much tricky having broad parametric
degeneracy and dealing with unknown DE.}

\begin{figure}
\mbox{\epsfxsize=8.2cm \epsffile{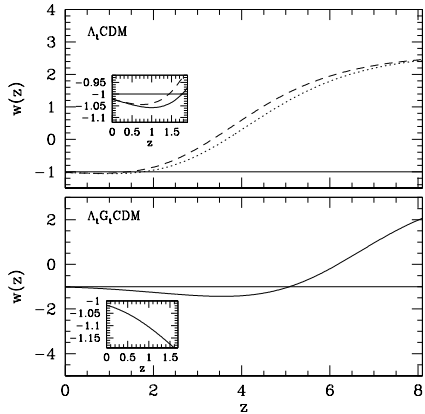}}
\caption{
{\em Upper Panel:} The $\CC_t$CDM evolution of the effective EoS parameter. We assume an improved
future accuracy $\Delta\Omega_m^0=-0.005$ and we consider two cases:
$\alpha=-\nu=3\times 10^{-3}$ (dashed line) and $\alpha=-3\nu/4$, with $\nu=-3\times 10^{-3}$ (dotted line).
{\em Lower Panel:} $w_{D}$ versus $z$ for the
$\CC_t$$G_t$CDM model under the same parameter inputs as with the previous model. In this case we have set $\alpha=0$
and we consider only the line for $\nu=-3\times 10^{-3}$.
{\em Inner Plots:}
The corresponding $w_{D}(z)$ for a more focused redshift interval near out time.}
\end{figure}

We have illustrated these features analyzing various types of dynamical
vacuum and/or evolving gravitational coupling $G$ models. These results
are timely in view of the recent features on the DE uncovered in the {\it
Planck} mission, which reinforce the possibility (persistently recorded in
the data releases by WMAP) of a phantom phase near our time. As it is
highly unlikely that this phase, if real at all, remained constant across
the cosmic evolution, we should suppose that it is a temporal one, and in
this sense it points to a possible dynamical evolution of the DE. However,
and most important, even if this phase would fade away from the next
generation of high precision cosmological experiments, the analysis
presented in this work shows that the usual parameterizations of the DE
may be too simpleminded to encompass the more realistic possibilities
offered by QFT, and in this sense we believe that it is unjustified to
conclude, on the basis of these parameterizations, that there is no
evidence of a dynamical vacuum energy. The observation of a fairly stable
vacuum at present (i.e. the successfulness of the $\CC$CDM) should not
lead us to conclude that this is the most natural expectation within the
context of fundamental physics. As a matter of fact, nothing should be
more natural in quantum field/string theory in an expanding universe that
a vacuum which is time evolving (hence redshifting) with the cosmic
expansion (Sol\`a 2011, 2013). A mild evolution at present would suggest a
more rapid evolution in the past, and this can be very useful to trace the
entire cosmic history in a framework where inflation, matter and dark
energy epochs can be encompassed in a unified framework (see Lima et al.
2013). We suspect that these are the kind of scenarios that should be able
to eventually explain the cosmological constant problem in its various
faces.

Let us finish by mentioning the fact that the recent observation, at the
CERN Large Hadron Collider, of a Higgs-like boson particle (Aad
\textit{et. al}, 2012; Chatrchyan \textit{et. al}, 2012) reminded us that
the vacuum energy (through the spontaneous symmetry breaking mechanism)
can be a fully tangible concept in real phenomenology. Despite the usual
difficulties associated with the value of the vacuum energy in QFT, the
palpable reality of the Higgs mechanism cannot be easily denied anymore,
and this means that we have to keep thinking on how to solve the CC
problem associated to the vacuum energy, not just blindly replacing it by
other concepts. Let us note that the difficulties with the CC problem are
no less severe than those associated to any ersatz entity trying to
substitute the CC. What is important is that the CC could be, after all, a
cosmic time evolving quantity. This is perfectly allowed by the
Cosmological Principle, and it could help in better dealing with the old
CC problem and the cosmic coincidence problem. In the meanwhile the
effects associated to changes in the vacuum energy can be as real as they
are in, say, the Casimir effect, which is sensitive not to the total
vacuum energy density but to changes in the vibrational modes of the
vacuum configuration. The possible analogy (Sol\`a, 2013) suggests that in
the cosmological case the changes in the vacuum energy density are
described in terms of mildly evolving functions of the expansion rate,
specifically in the form of powers of $H^2$ and $\dot{H}$.

{To summarize, in this work we have emphasized on the following two
important issues. First, there are some theoretical models that while they
do not have any theoretical ghost, their effective equation of state can
have a phantom like behavior and such models can fit the data pretty well
comparable with cosmological constant. Second, parametric degeneracies
(or, as it called also, cosmographic degeneracies) along with unknown
nature of dark energy makes it very tricky to work with parametric forms
of the equation of state of dark energy fitting cosmological data.}

We have shown that these effects could potentially be detected in future
observations through the dynamical features encoded in the effective
equation of state of the dark energy. We have also emphasized that we
should  stay open minded to this possibility. Although the usual
parameterizations of the EoS existing in the literature might not be able
to capture these effects at present, we expect that future high precision
observations and the use of more general parameterizations should help in
eventually unraveling the dynamical nature of the cosmic vacuum energy.

\section*{Acknowledgments}
We thank V. Sahni for interesting comments on our work and for pointing
out some relevant references. SB acknowledges support by the Research
Center for Astronomy of the Academy of Athens in the context of the
program {\it ``Tracing the Cosmic Acceleration''}. JS has been supported
in part by projects FPA2010-20807 and CPAN (Consolider CSD2007-00042).
Also by 2009SGR502 Generalitat de Catalunya. SB thanks the Dept.\ ECM of
the Univ.\ de Barcelona for the hospitality when part of this work was
being done.

\newcommand{\JHEP}[3]{ {JHEP} {#1} (#2)  {#3}}
\newcommand{\NPB}[3]{{ Nucl. Phys. } {\bf B#1} (#2)  {#3}}
\newcommand{\NPPS}[3]{{ Nucl. Phys. Proc. Supp. } {\bf #1} (#2)  {#3}}
\newcommand{\PRD}[3]{{ Phys. Rev. } {\bf D#1} (#2)   {#3}}
\newcommand{\PLB}[3]{{ Phys. Lett. } {\bf B#1} (#2)  {#3}}
\newcommand{\EPJ}[3]{{ Eur. Phys. J } {\bf C#1} (#2)  {#3}}
\newcommand{\PR}[3]{{ Phys. Rep. } {\bf #1} (#2)  {#3}}
\newcommand{\RMP}[3]{{ Rev. Mod. Phys. } {\bf #1} (#2)  {#3}}
\newcommand{\IJMP}[3]{{ Int. J. of Mod. Phys. } {\bf #1} (#2)  {#3}}
\newcommand{\PRL}[3]{{ Phys. Rev. Lett. } {\bf #1} (#2) {#3}}
\newcommand{\ZFP}[3]{{ Zeitsch. f. Physik } {\bf C#1} (#2)  {#3}}
\newcommand{\MPLA}[3]{{ Mod. Phys. Lett. } {\bf A#1} (#2) {#3}}
\newcommand{\CQG}[3]{{ Class. Quant. Grav. } {\bf #1} (#2) {#3}}
\newcommand{\JCAP}[3]{{ JCAP} {\bf#1} (#2)  {#3}}
\newcommand{\APJ}[3]{{ Astrophys. J. } {\bf #1} (#2)  {#3}}
\newcommand{\AMJ}[3]{{ Astronom. J. } {\bf #1} (#2)  {#3}}
\newcommand{\APP}[3]{{ Astropart. Phys. } {\bf #1} (#2)  {#3}}
\newcommand{\AAP}[3]{{ Astron. Astrophys. } {\bf #1} (#2)  {#3}}
\newcommand{\MNRAS}[3]{{ Mon. Not. Roy. Astron. Soc.} {\bf #1} (#2)  {#3}}
\newcommand{\JPA}[3]{{ J. Phys. A: Math. Theor.} {\bf #1} (#2)  {#3}}
\newcommand{\ProgS}[3]{{ Prog. Theor. Phys. Supp.} {\bf #1} (#2)  {#3}}
\newcommand{\APJS}[3]{{ Astrophys. J. Supl.} {\bf #1} (#2)  {#3}}

\newcommand{\Prog}[3]{{ Prog. Theor. Phys.} {\bf #1}  (#2) {#3}}
\newcommand{\IJMPA}[3]{{ Int. J. of Mod. Phys. A} {\bf #1}  {(#2)} {#3}}
\newcommand{\IJMPD}[3]{{ Int. J. of Mod. Phys. D} {\bf #1}  {(#2)} {#3}}
\newcommand{\GRG}[3]{{ Gen. Rel. Grav.} {\bf #1}  {(#2)} {#3}}



\end{document}